\documentclass[aps,prb,twocolumn,groupedaddress,showpacs,floatfix,amsmath]{revtex4}
\def\bc{\begin{center}}
\def\ec{\end{center}}
\def\be{\begin{equation}}
\def\ee{\end{equation}}
\def\bw{\begin{widetext}}
\def\ew{\end{widetext}}

\usepackage{graphicx}
\usepackage{dcolumn}
\usepackage{bm}
\begin{document}
\title{Electron operator at the edge of the 1/3 fractional quantum Hall liquid}
\author{ Shivakumar Jolad, Chia-Chen Chang, and Jainendra K. Jain }
\affiliation{Department of Physics, Pennsylvania State University, University Park, PA 16802}\date{\today}

\begin{abstract}
This study builds upon the work of Palacios and MacDonald 
(Phys. Rev. Lett.  {\bf 76}, 118 (1996)), wherein they identify the bosonic excitations 
of Wen's approach for the edge of the 1/3 fractional quantum Hall state with certain 
operators introduced by Stone.  Using a quantum Monte Carlo method, we extend to 
larger systems containing up to 40 electrons and obtain more accurate 
thermodynamic limits for various matrix elements for a short range interaction.  
The results are in agreement with those of Palacios and MacDonald for small systems, but offer further insight into the detailed approach to the thermodynamic limit.  For the short 
range interaction, the results are consistent with the chiral Luttinger liquid predictions.
We also study excitations using the Coulomb ground state for up to nine electrons
to ascertain the effect of interactions on the results; in this case our tests of the chiral 
Luttinger liquid approach are inconclusive.  
\end{abstract}

\maketitle

\section{Introduction}

The low-energy excitations of an ordinary Landau Fermi liquid
resemble electrons, and are perturbatively accessible from the free system. 
In recent years, there has been much interest in systems that do not conform to 
the Landau Fermi liquid paradigm.  An example of a non-Landau Fermi liquid (NLFL) is 
the system of interacting fermions in one dimension, called a 
Tomonaga-Luttinger  liquid, which is described by nonperturbative techniques 
such as bosonization [\onlinecite{GiulianiVignale}].  For Landau-Fermi liquids the electron spectral function 
has a sharp peak with a nonzero weight, called the quasiparticle renormalization 
factor.  The quasiparticle renormalization factor vanishes for interacting 
electrons in one dimension.

For a FQHE state [\onlinecite{Tsui}], the excitations in the bulk are gapped, but
arbirtarily low-energy 
excitations are available at the edge.  The edge dynamics is one dimensional, and 
constitutes a realization of a chiral Luttinger liquid, which is a Luttinger liquid 
consisting of fermions moving only in one direction [\onlinecite{WenIntJModPhy, Chang}].  Here, only electrons at one edge of 
the FQHE system are considered; coupling with the oppositely moving electrons at 
the other edge is neglected, which is a good approximation for wide samples 
for which the two edges are spatially far separated.  Wen has proposed an 
effective chiral Luttinger liquid (ECLL) model 
for the description of the FQHE edge [\onlinecite{WenIntJModPhy}], 
according to which the long distance properties at the edge are universal, 
described by a quantized exponent the value of which is determined 
by the quantized Hall resistance of the bulk FQHE state.  
The most direct probe of the properties of this liquid is 
tunneling of an external electron laterally into the edge of the 
FQHE system.  Non-linear I-V characteristics 
have demonstrated NLFL behavior for the FQHE edge.  The ECLL approach
predicts an I-V behavior for the tunnel conductivity of the form $I\propto V^\alpha$, 
with a universal value for $\alpha$.  In particular, for FQHE states at 
filling factors $\nu=n/(2pn+1)$, the effective approach predicts $\alpha=2p+1$.
Experiments find a nontrivial value for $\alpha$ (i.e. $\alpha\neq 1$), but 
the observed $\alpha$ deviates from the predicted one [\onlinecite{Grayson, Chang1,Chang2}].  This discrepancy 
has motivated much work [\onlinecite{Chang,Conti,MandalJain,ZulPalMacD,Lopez,Shytov}],
including the present paper.

The ECLL  approach is built upon the idea of a one-to-one correspondence between 
the fermionic and bosonic Fock spaces in one dimension, and identifies 
a relationship between the operators of the two problems.  In particular, the 
fermionic field operator $\hat\psi(x)$ is related to the bosonic field operator 
$\hat\phi(x)$ through 
the expression $\hat\psi(x)\sim \exp[-i\hat\phi(x)]$, which can be established 
rigorously for the ordinary one dimensional systems [\onlinecite{GiulianiVignale}].  A similar rigorous derivation  has not been possible for the electron field operator at the edge of a FQHE 
system.   Wen postulates that the electron operator at the edge of the $\nu=1/m$ FQHE  
system is given by 
\begin{equation}
\hat\psi(x)\sim e^{-i\sqrt{m}\hat\phi(x)}, 
\label{Wenansatz}
\end{equation}
which has the virtue of 
satisfying the antisymmetry property when $m=2p+1$ is an odd integer [\onlinecite{WenIntJModPhy}].  This 
form leads to the quantized exponent for the I-V of the tunnel conductance.  It is 
not known at the present what causes the discrepancy between the effective 
theory and experiment.  

Our work is an outgrowth of the exact diagonalization studies of Palacios and 
MacDonald [\onlinecite{PalaciosMacDonald}] at $\nu=1/3$, the results of which were 
interpreted by the authors as confirming Wen's 
ansatz.  We now know, however, that the discrepancy is small at $\nu=1/3$ 
(the observed value of $\alpha 
\approx 2.7$ is close to the predicted $\alpha=3$), and the results of Ref. [\onlinecite{PalaciosMacDonald}] were not sufficiently accurate to capture such small deviations. 
We extend their calculations to larger systems, using a Monte Carlo method, to 
obtain more accurate thermodynamic extrapolations. 
Furthermore, Ref. [\onlinecite{PalaciosMacDonald}] assumed a short range interaction 
model; we also investigate to what extent the results are sensitive to the 
form of the interaction.  Our results are consistent with 
the ECLL predictions for the short-range model, but inconclusive for the Coulomb interaction.

It should be stressed that the FQHE liquid itself provides a beautiful paradigm for 
a breakdown of the Landau-Fermi liquid concept.  Here, strong interactions generate electron-vortex bound 
states called composite fermions, which are qualitatively distinct from, and 
perturbatively unrelated to, electrons [\onlinecite{JainCFpaper,JainCFBook}].   Composite fermions experience a greatly 
reduced effective magnetic field, and possess quantum numbers (for the local 
charge and braiding statistics) which are a fraction of the electron quantum numbers.
The bulk properties of the composite fermion (CF) liquid have been investigated by a variety of means, 
and many experiments have directly verified the effective magnetic field, a 
clear indication of the NLFL nature of the state. 
The question of how an external electron, tunneling vertically in the CF liquid,
couples to the system has been studied, and it has been predicted that 
it tunnels resonantly into an excited state, which is a bound state of several excited composite fermions, to produce a sharp peak in the electron spectral function [\onlinecite{JainPet, Vignale}].  The CF liquid therefore provides a different mechanism for the 
breakdown of the Landau-Fermi liquid concept:  The electron renormalization factor 
remains nonzero, but lower energy states appear that are not described 
in terms of new quasiparticles.

The plan of the paper is as follows.  Section II lists the connection between the bosonic approach and wave functions.  The results are given in Sec. III, followed by conclusion 
in Sec. IV.

\section{Spectral weights}

Following Palacios and 
MacDonald [\onlinecite{PalaciosMacDonald}] we test the validity of Eq. \ref{Wenansatz} 
by comparing the microscopically calculated spectral weights 
\be
C_{\{n_l\}}= \frac{\langle \{n_l\} | \hat\psi^\dagger(\theta)|0\rangle}{\langle 0 | \hat\psi^\dagger(\theta)|0\rangle}
\label{spectralweight}
\ee
(where $\theta$ denotes the position in one dimension, wrapped into a circle via 
periodic boundary conditions) 
with the predictions of the ECLL  approach.  In the ECLL approach  
at filling factor $1/m$ [\onlinecite{WenIntJModPhy}], the vacuum state 
$|0\rangle$ contains no bosons and the various 
symbols have the following meaning:
\begin{equation}
\hat{\psi}^\dagger (\theta)=
\sqrt{z} e^{i\sqrt{m}\hat{\phi}_+(\theta)}e^{-i\sqrt{m}\hat{\phi}_-(\theta)}
\label{FieldOp}
\end{equation}
\be
\hat{\phi}_+(\theta)=-i\sum_l \frac{1}{\sqrt{l}} a_l^\dagger e^{il\theta}
\ee
\be
\hat{\phi}_-(\theta)=-i\sum_l \frac{1}{\sqrt{l}}a_l e^{-il\theta}
\ee
and
\be
|\{n_l\}\rangle = \prod_{l=0}^\infty \frac{\left(a_l^{\dagger}\right)^{n_l}}{\sqrt{n_l!}}
|0\rangle.
\ee 
Here $a_l^\dagger$ and $a_l$ are creation and annihilation operators for a boson  
in the angular momentum $l$ state, with the total angular momentum given by
\be
\Delta M= \sum_l  l \,n_l .
\ee
By expanding  $\exp[i\sqrt{m}\hat{\phi}_+(\theta)]$ and 
$\exp[-i\sqrt{m}\hat{\phi}_-(\theta)]$ in power series of products of  $a_l$ and $a_l^\dagger$, it is straightforward to compute the analytical values for the spectral weights 
$C_{\{n_l\}}$.  The square of the spectral weight $|C_{\{n_l\}}|^2$ is independent of the angular position parameter. The denominator in Eq. \ref{spectralweight} eliminates the unknown normalization constant $\sqrt{z}$ in Eq. \ref{FieldOp}.

From the perspective of electrons, the vacuum state is the ground state of 
interacting electrons at $\nu=1/m$, and the field operator has the standard meaning 
of
\be
\hat\psi^\dagger(\theta)=\sum_l\eta_l^*(\theta) c_l^\dagger\equiv \sum_l \psi_l^\dagger(\theta) ,
\ee
where $c_l^\dagger$ and $c_l$ are creation and annihilation operators for an electron 
in the angular momentum $l$ state, the wave function for which is $\eta_l$:
\be
\eta_l(z)=\frac{z^l}{\sqrt{2\pi 2^l l!}} e^{-|z|^2/4},\;\;z=x-iy.
\ee
The denominator of Eq. \ref{spectralweight} is 
interpreted as 
\be
\langle 0|\hat\psi^\dagger(\theta)|0\rangle = \langle \Psi_0^{N+1}|\hat\psi^\dagger(\theta)|\Psi_0^N\rangle
\ee
where $\Psi_0^N$ is the normalized ground state of $N$ interacting electrons at $\nu=1/m$.
The numerator is interpreted as
\be
\langle {\{n_l\}} |\hat\psi^\dagger(\theta)|0\rangle = \langle \Psi_{\{n_l\}}^{N+1}|\hat\psi_L^\dagger(\theta)|\Psi_0^N\rangle.
\label{eq9}
\ee
Here, we have,
\begin{equation}
\label{ElectronCreate}
\hat\psi_L^\dagger |\Psi_{0}^N\rangle=\mathcal{N}_L\mathcal{A}\left[ z_{N+1}^Le^{-|z_{N+1}|^2/4}\Psi_0^N(z_1,z_2\dots,z_N)\right],
\end{equation}
where ${\cal A}$ is the antisymmetrization operator and $\cal{N}_L$ is the normalization constant.

The wave function $\Psi_{\{n_l\}}^{N+1}$, the electronic counterpart of the bosonic state $|\{n_l\}\rangle$, represents an excited state that involves increasing the angular momentum of the $N$ particle ground state by $L$ units.  It is 
not immediately obvious how to construct it for a general case.  For the 
integral quantum Hall state at $\nu=1$, Stone showed [\onlinecite{stone1,stone2, Oaknin}] that it is obtained 
by multiplying the ($N+1$ particle) ground state by the factor $\prod_l S_l^{n_l}$,
where $S_l$ are defined as:
\begin{equation}
\label{StoneOpDef}
S_l=\sum_{j=1}^N z_j^l.
\end{equation}
The product increases the total angular momentum of the $N+1$ particle 
ground state by 
$\Delta M=\sum_l l n_l $.
The composite-fermion analogy suggests the identification  [\onlinecite{PalaciosMacDonald}]
\be
|\Psi^{N+1}_{\{n_l\}}\rangle = \mathcal{N}_{n_l}\left(\prod_l (S_l)^{n_l}\right) \Psi_0^N 
\label{identification}
\ee
at $\nu=1/m$, where $\mathcal{N}_{n_l}$ is the normalization constant.  The angular momentum changes $L$ and $\Delta M$, 
defined relative to the $N$ and $N+1$ particle ground states, respectively, are related by
\be
L=\Delta M+mN ,
\ee
where the last term is the difference between the angular momenta of the 
ground states.
We will assume the identification in Eq. \ref{identification} in what follows;
further justification for it is given in the Appendix.  In our calculations with 
exact diagonalization method, we will use that the 
second quantization representation for the operators $S_l$
is given by (apart from a constant factor)
\begin{equation}
S_l=\sum_{k=0}^{\infty}\sqrt{\frac{(k+l)!}{k!}}c_{k+l}^\dagger c_k.
\label{S_operator}
\end{equation}

\section{Results}

\subsection{Exact diagonalization: Short range interaction}

Palacios and MacDonald [\onlinecite{PalaciosMacDonald}] computed the squared spectral 
weights $|C_{\{n_l\}}|^2$ defined in Eq. \ref{spectralweight} by exact diagonalization for a short-range 
interaction model for which the Laughlin wave function [\onlinecite{Laughlin}] 
and  the excited states in Eq. \ref{identification} are exact eigenstates. (This interaction 
takes a nonzero value for the $V_1$ pseudopotential [\onlinecite{HaldanePseud}] but 
sets all other pseudopotentials to zero.) They obtain results for systems containing 
up to eight electrons for $\Delta M=$1 to 4; from an extrapolation 
to the thermodynamic limit, they find approximate consistency 
with the predictions of the effective ECLL  approach.

The results  we obtained for $V_1$ interaction are shown in Table \ref{table1}, and are identical to the results by Palacios and MacDonald  [\onlinecite{PalaciosMacDonald}]. 
However, our extrapolation to thermodynamic limit differs slightly from theirs. The 
extrapolation assumes
a leading finite size correction for $|C_{\{n_l\}}|^2$ proportional to $1/N$; 
this dependence has not been derived analytically, but matches the numerical data  
well for these particle numbers.  
The thermodynamic values are close to, but significantly different from,  
those predicted by the ECLL  model.

\subsection{``Exact" diagonalization: Coulomb interaction}
A proper extension of the above results to the Coulomb interaction is not known.  In the  
above, both the ground and excited states are exact eigenstates of the $V_1$ model.  While the exact ground state for the Coulomb interaction can be obtained for small systems, no operators analogous to the $S_l$ of Eq. 
\ref{StoneOpDef} are known that produce exact 
excited states.  We use an approximate, ``hybrid" approach, in which we take the exact Coulomb ground state, but use the same operators 
$S_l$ to create excited states.  The above calculation can then be extended to the Coulomb interaction.  The exact spectral weights with for angular momenta $\Delta M=1$ to 4 and particles $N=$4 to 9 are tabulated in Table \ref{table2}.  Again, there is a small, but significant, deviation from the ECLL  results.

We mention here a technical point relating to the efficiency of the numerical calculation.
The number of basis vectors increases rapidly with the number of particles.
For example, for $ \Delta M=0, N= 8$ and $9$ and $10$, the number of basis  vectors  are 55,974, 403,016 and 2,977,866, respectively.   During bosonic state creation and overlap calculation, the most time consuming step is searching for a given component from the ket vector $\psi^{\dagger}_{L}|\Psi_0^N \rangle$ and matching it with the corresponding bra vector component $\langle \Psi_{\{n_l\}}^{N+1} |$.   
We collect the fermionic states $|m_1,m_2,\dots, m_{N}\rangle $ into bins indexed by the smallest three angular momentum values $\{m_1,m_2,m_3\}$, which allows the search for a match to be restricted to a single bin.  This reduces the computation time by a factor of 70, enabling computation of single boson states within 35 hours on our supercomputing cluster.

\bw

\begin{table}[t]
\label{ExDiag$V_1$}
\begin{tabular}{ccccccccc}\hline\hline
$\Delta M$ & $\{n_l\}$  &  $N=4$   &  $N=5$  &  $N=6$  &  $N=7$  &   $N=8$ &N$\rightarrow\infty$  & ECLL \\\hline
      1    & $\{1000\}$ &  2.6000  &  2.6667 &  2.7142 & 2.7500&2.778  &2.9532 &3   \\
      2    & $\{2000\}$ &  3.6400  &  3.7778 &  3.8755 & 3.9531&4.012  & 4.3769 &9/2 \\
           & $\{0100\}$ &  1.2585  &  1.2953 &  1.3224 & 1.3425 &1.358 & 1.4562 &3/2 \\
      3   & $\{3000\}$ &  3.6400  &  3.7778 &  3.8775 & 3.9531&4.012  & 4.3777 &9/2 \\
           & $\{1100\}$ &  3.6740  &  3.8133 &  3.9128 &  3.9860&4.041   & 4.4053 &9/2 \\
           & $\{0010\}$ &  0.9356  &  0.9358 &  0.9390 & 0.9425 &0.946 & 0.9543 &1   \\
    4    & $\{4000\}$ &  2.9120  &  2.9907 &  3.0466 &3.088&3.121  & 3.3267 &27/8\\
           & $\{2100\}$ &  5.6503  &  5.8730 &  6.0247 &  6.131&6.209 & 6.7701 &27/4\\
           & $\{1010\}$ &  2.7489  &  2.7952 &  2.8284 &   2.852&2.869 &  2.9894 &3   \\
           & $\{0200\}$ &  1.0184  &  1.0353 &  1.0484 &  1.058&1.064 & 1.1102 &9/8 \\
           & $\{0001\}$ &  0.9044  &  0.8583 &  0.8302 & 0.8109&0.797  & 0.6881 &3/4 \\
\hline\hline
\end{tabular}
\caption{Squared spectral weights $|C_{\{n_l\}}|^2$ from exact diagonalization method for the short range model in which only the $V_1$ pseudopotential 
is nonzero.  $N$ is the number of electrons in the study.
For finite $N$, our numbers are identical to those in 
Ref. \onlinecite{PalaciosMacDonald}, but our extrapolations to $N\rightarrow\infty$ 
are slightly different. "(A linear extrapolation in $1/N$ is assumed for obtaining the thermodynamic limits in this table, in Tables II and III, and in Figs. 1 and 2 for the exact diagonalization results;  the validity of this assumption is questioned by larger system Monte Carlo results shown in Fig. 1 for the short-range interaction.)
\label{table1}} 
\end{table}

\begin{center}
\begin{table}[t]
\label{EDCoulTable}
\begin{tabular}{ccccccccc	c}\hline\hline
$\Delta M$ & $\{n_l\}$  &  $N=4$   &  $N=5$  &  $N=6$  &  $N=7$  &   $N=8$ &   $N=9$ &N$\rightarrow\infty$& ECLL \\\hline
      1    & $\{1000\}$ &  2.6000  &  2.6670 &  2.7150 & 2.7514  &2.7801&2.8031& 2.9619 &3  \\
      2    & $\{2000\}$ &  3.6400  &  3.7783 &  3.8786 &3.9553  &4.0157&4.0644& 4.3954 &9/2 \\
           & $\{0100\}$ &  1.2871  &  1.3177 &  1.3504 & 1.3654  &1.3744& 1.3849& 1.4658 &3/2 \\\
      3    & $\{3000\}$ &  3.6400  &  3.7783 &  3.8786 & 3.9554  &4.0157& 4.0644& 4.3954 &9/2 \\
           & $\{1100\}$ &  3.7551  &  3.8768 &  3.9937 & 4.0522  & 4.0885& 4.1265& 4.4334 &9/2\\
           & $\{0010\}$ &  1.0029  &  0.9836 &  0.9925 & 0.9830  &0.9780&0.9773& 0.9588  &1\\
      4    & $\{4000\}$ &  2.9120  &  2.9911 &  3.0475 & 3.0903  &3.1233&3.1499 & 3.3365& 27/8\\
           & $\{2100\}$ &  5.7699  &  5.9657 &  6.1450 &6.2311   & 6.2804&  6.3315& 6.8044 &27/4\\
           & $\{1010\}$ &  2.9313  &  2.9315 &  2.9843 & 2.9710  &2.9636&2.9662 &   3.0087 &3\\
           & $\{0200\}$ &  1.0632  &  1.0700 &  1.0926 & 1.0935  &1.0886&1.0898  & 1.1193& 9/8\\
           & $\{0001\}$ &  0.9676  &  0.9128 &  0.9095 &0.8725   &0.8420& 0.8264 & 0.7259 & 3/4\\
\hline\hline
\end{tabular}
\caption{Squared spectral weights  $|C_{\{n_l\}}|^2$ obtained from exact diagonalization  method for the Coulomb interaction.\label{table2}}
\end{table}
\end{center}
\ew

\subsection{Monte Carlo simulation results}

The exact diagonalization method 
has the drawback of being restricted to small numbers of particles.  Fortunately with quantum Monte Carlo methods we can extend the results to much larger systems for the 
above-mentioned short-ranged $V_1$ interaction, for which all the wave functions in 
question are explicitly known.  With the Monte Carlo calculation method we have 
obtained the spectral weights for up to 40 particles, for $\Delta M$ ranging from 1 to 6. The results are shown in Table \ref{table3} and  Figs. \ref{EDandMC} and \ref{MCM5M6}.  

\subsection{Orthogonality}

 The ECLL  approach also predicts orthogonality between states $|\Psi_{\{n_l\}}\rangle$ and $|\Psi_{\{n_l'\}}\rangle$ for $\{n_l\}\ne \{n_l'\}$. Such an orthogonality is not apparent from the wave function, and does not follow from any symmetry. We have 
numerically tested it for several cases; the results, summarized
in Table \ref{table4}, indicate that  the overlap between different states rapidly diminishes with increasing number of particles. This also demonstrates that the states generated by the operators $S_l$ 
do not exactly represent free boson states for finite $N$, but are meaningful only in the thermodynamic limit.

\section{Discussion and conclusions}

The thermodynamic values of the spectral weights, obtained by assuming a 
linear fit as a function of $1/N$, lie within a few percent of the ECLL  predictions
(0.9 \% to 18 \%  for single boson states; see Table \ref{table3}), but 
the deviations between the two are often significant.  However,  deviations 
from linear fit are seen for large $N$.
For example,  for $\{1000\}, \{0100\}, \{3000\}, \{1100\}, \{0100\}, \{0001\},\{010100\}$, 
$\{00001\},\{200100\}, \{1000010\},\{01010\}, \{002000\} $, $\{000001\}$ (Figs. 
\ref{EDandMC} and \ref{MCM5M6}), the actual fit is nonlinear, and tends toward the ECLL  results with increasing $N$.  
 A proper thermodynamic value is difficult to estimate accurately in many cases, partly  
because of the lack of analytic results regarding the $N$ dependence of $|C_{\{n_l\}}|^2$.  However, the thermodynamic limits are in general agreement with the ECLL  
predictions.  This is consistent with several previous studies [\onlinecite{MandalJain,ZulPalMacD,Shytov}] that confirm  
the validity of the ECLL model for the short range interaction.

With a partial inclusion of the Coulomb interaction, linear extrapolation of the results 
for up to nine particles again gives matrix elements that are close to the ECLL prediction 
but significantly different.  The results are also in general quite different from those 
for short range interaction.  Here, however, it is not possible to extend our 
study to larger systems.  From our experience with the short range interaction, 
we cannot rule out deviation of the $|C_{\{n_l\}}|^2$ vs. $1/N$ plot 
from linearity for large $N$, and therefore consider our study as being inconclusive. 

Our calculations essentially serve as a 
test of the identification of the operators $S_l$ with the boson operators $a_l$ 
for the short range interaction.  
This was first introduced by Palacios and MacDonald, and is further justified 
in the appendix, but not proven rigorously.

\section{Acknowledgement}
One of us (S.J.) acknowledges Paul Lammert for many discussions and thanks 
Gun Sang Jeon and Csaba T\"oke for help with programming algorithm. We are grateful to 
A. H. MacDonald and J. J. Palacios for useful communications.  
The computational work was performed on the Lion-XO cluster of High Performance Computing Group, Pennsylvania State University.  Partial support by the National 
Science Foundation under grant no. DMR-0240458 is acknowledged.
\begin{widetext}

\begin{table}[htbp]
\label{MCTable}
\begin{tabular}{clclc}\hline\hline
$\Delta M$ &$\{n_l\} $& $|C_{\{n_l\}}|^2$&\quad $|C^{ECLL }_{\{n_l\}}|^2$ & \% deviation \\
\hline
1&    \{1000\}&		2.9732(0.02)&\quad 	3   &			0.89 \\
2&	\{2000\}&		4.5136(0.02)&\quad 	4.5&			0.30\\
   &    \{0100\}  &	1.4760(0.01)&\quad 	1.5&			1.60\\
3&	\{3000\}&		4.4896(0.02)&\quad 	4.5&			0.23\\
&	\{1100\}&		4.4764(0.02)&\quad 	4.5&			0.52\\
&	\{0010\}&		0.9769(0.01)&\quad 	1&			2.31\\
4&	\{4000\}&		3.3833(0.03)&\quad 	3.375&		0.25\\
&	\{2100\}&		6.7928(0.05)&\quad 	6.75	&		0.63\\
&	\{1010\}&		3.0005(0.02)&\quad 	3&			0.02\\
&	\{0200\}&		1.1273(0.01)&\quad 	1.125&		0.20\\
&	\{0001\}&		0.7210(0.01)&\quad 	0.75	&		3.87\\
5&	\{50000\}&	2.0429(0.02)&\quad 	2.025&		0.88\\
&	\{31000\}&	6.9042(0.04)&\quad 	6.75	&		2.28\\
&	\{20100\}&	4.5869(0.03)&\quad 	4.5	&		1.93\\
&	\{10010\}&	2.2493(0.02)&\quad 	2.25	&		0.03\\
&	\{01100\}&	1.5073(0.01)&\quad 	1.5	&		0.49\\
&	\{00001\}&	0.5481(0.01)&\quad 	0.6	&		8.65\\
6&	\{600000\}&	1.0303(0.01)&\quad 	1.0125 &		1.76\\
&	\{410000\}&	5.2688(0.04)&\quad 	5.0625&		4.08\\
&	\{301000\}&	4.6980(0.04)&\quad 	4.5	&		4.40\\
&	\{200100\}&	3.5203(0.03)&\quad 	3.375 &		4.31\\
&	\{220000\}&	5.2251(0.04)&\quad 	5.0625 &		3.21\\
&	\{100010\}&	1.7837(0.02)&\quad 	1.8	&		0.91\\
&	\{111000\}&	4.7573(0.03)&\quad 	4.5	&		5.72\\
&	\{010100\}&	1.1669(0.01)&\quad 	1.125 & 		3.72\\
&	\{002000\}&	0.5014(0.01)&\quad 	0.5	&		0.28\\
&	\{000001\}&	0.4072(0.01)&\quad 	0.5  	&		18.56\\
\hline\hline
\end{tabular}
\caption{Monte Carlo simulation results  for $V_1$ pseudopotential with 20 particles. Third column shows the thermodynamic extrapolation of squared spectral weight. The statistical errors are shown in parenthesis. ECLL  predictions and deviation from ECLL  predictions are shown in fourth and fifth columns, respectively.\label{table3}}
\end{table}

\begin{table}[h]
\label{OrthTestTable}
\begin{tabular}{cccccc}\hline\hline
$\Delta M$ & $\{n_l\}$ &$\{n_l'\}$  &  $N=6$  \quad &  $N=8$ \quad &  $N=15$  \\ \hline
2 &    $\{2000\}$&$\{0100\}$&    6.90(0.1)            $\times 10^{-3}$\quad&	4.3(0.12)           $\times 10^{-3}$\quad&	1.3(0.06)              $\times 10^{-3}$ \\
3&	$\{3000\}$&$\{1100\}$&	2.1(0.05)          $\times 10^{-2}$  \quad&	1.2(0.03)           $\times 10^{-2}$\quad&	4.0(0.06)               $\times 10^{-3}$ \\
 &	$\{3000\}$&$\{0010\}$&	1.0(0.07)            $\times 10^{-4}$\quad&      8.6(0.52)	         $\times 10^{-5}$\quad&	1.2(0.57)               $\times 10^{-5}$ \\
&	$\{1100\}$&$\{0010\}$&	4.3(0.03)            $\times 10^{-2}$\quad&	2.7(0.10)	         $\times 10^{-2}$\quad&	7.7(0.24)               $\times 10^{-3}$\\
4&	$\{4000\}$&$\{2100\}$&	4.0(0.05)           $\times 10^{-2}$\quad&	2.5(0.03)		$\times 10^{-2}$\quad&	8.1(0.20)		   $\times 10^{-3}$ \\
&	$\{4000\}$&$\{1010\}$&	4.20(0.2)            $\times 10^{-4}$\quad&	1.5(0.20)		$\times 10^{-4}$\quad&	4.0(1.00)		   $\times 10^{-5}$  \\
&	$\{4000\}$&$\{0200\}$&	1.0(0.05)	         $\times 10^{-4}$\quad&	5.0(0.30)		$\times 10^{-5}$\quad&	5.5(1.40)		   $\times 10^{-6}$ \\
&	$\{4000\}$&$\{0001\}$&	4.20(1.2)	          $\times 10^{-6}$\quad&	1.0(0.2.4)		$\times 10^{-6}$\quad&	2.7(1.20)	            $\times 10^{-6}$  \\
&	$\{2100\}$&$\{1010\}$&	8.1(0.05)        	$\times 10^{-2}$\quad&	5.0(0.04)		$\times 10^{-2}$\quad&	1.6(0.01)		   $\times 10^{-2}$  \\
&	$\{2100\}$&$\{0200\}$&	1.4(0.02)	        $\times 10^{-2}$\quad&	8.3(0.80)		$\times 10^{-3}$\quad&	2.7(0.10)		   $\times 10^{-3}$  \\
&	$\{2100\}$&$\{0001\}$&	1.8(0.05)	         $\times 10^{-3}$\quad&	6.8(0.30)		$\times 10^{-4}$\quad&	8.0(1.10) 		   $\times 10^{-5}$\\
&	$\{1010\}$&$\{0200\}$&	1.1(0.03)	         $\times 10^{-3}$\quad&	4.0(0.10)		$\times 10^{-4}$\quad&	4.2(0.90)		   $\times 10^{-5}$ \\
&	$\{1010\}$&$\{0001\}$&	8.8(0.05)	          $\times 10^{-2}$\quad&	5.3(0.04) 		$\times 10^{-2}$\quad&	1.6(0.02)		   $\times 10^{-2}$ \\
&	$\{0200\}$&$\{0001\}$&	6.0(0.04)	      	$\times 10^{-2}$\quad&	3.5(0.03)		$\times 10^{-2}$\quad&	1.0(0.03)		   $\times 10^{-2}$ \\
\hline\hline
\end{tabular}
\caption{Overlaps $|\langle \Psi_{\{n_l\}} |\Psi_{\{n_l'\}}\rangle|^2$, where $\Psi_{\{n_l\}}$ is defined as in Eq. (\ref{identification}). Results are shown for $\Delta$M=1 to 4 
for several values of $N$.\label{table4}}
\end{table}

\begin{center}
\begin{figure}
\includegraphics[viewport=60 250 550 675]{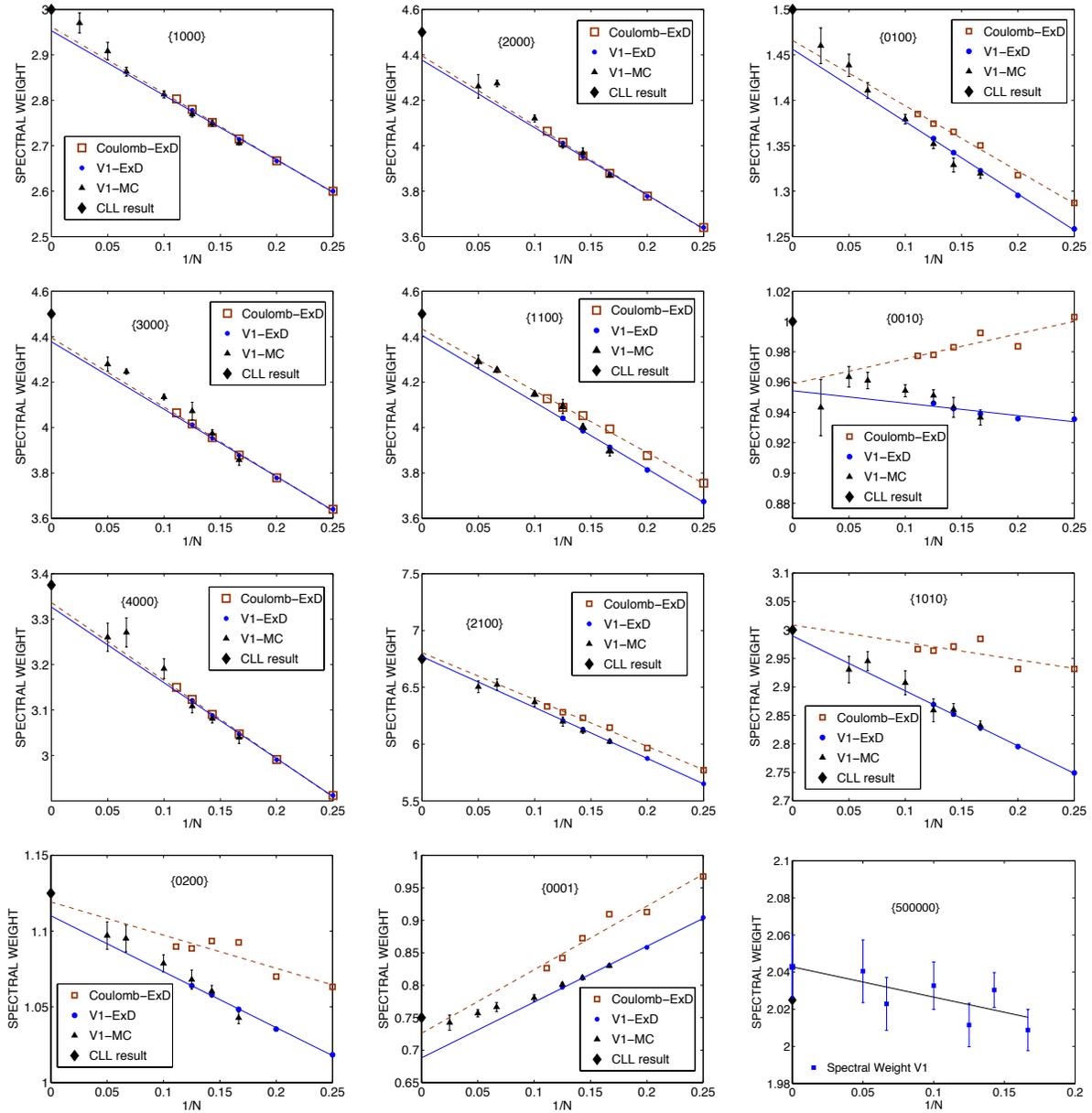}
\caption{(Color online) Squared spectral weights $|C_{\{n_l\}}|^2$ obtained from several methods: exact diagonalization for short range interaction (labeled V1-ExD) and Coulomb interaction (Coulomb-ExD), and Monte Carlo evaluation for the short range interaction (V1-MC).  The symbols are defined on the figures.  The error bars indicate the statistical uncertainty in the Monte Carlo results. 
Results are shown for $\Delta M=1,2,3,4$ and 5. For $\Delta M=5$, $\{50000\}$ state we have only Monte Carlo results. The solid line and the dashed 
line are a linear fit for the exact diagonalization results for short range and Coulomb 
interactions, respectively.  The chiral Luttinger liquid (CLL) predictions are marked by solid diamonds on the vertical axes.}
\label{EDandMC}
\end{figure}
\end{center}

\begin{figure}
\includegraphics[viewport=60 150 550 725]{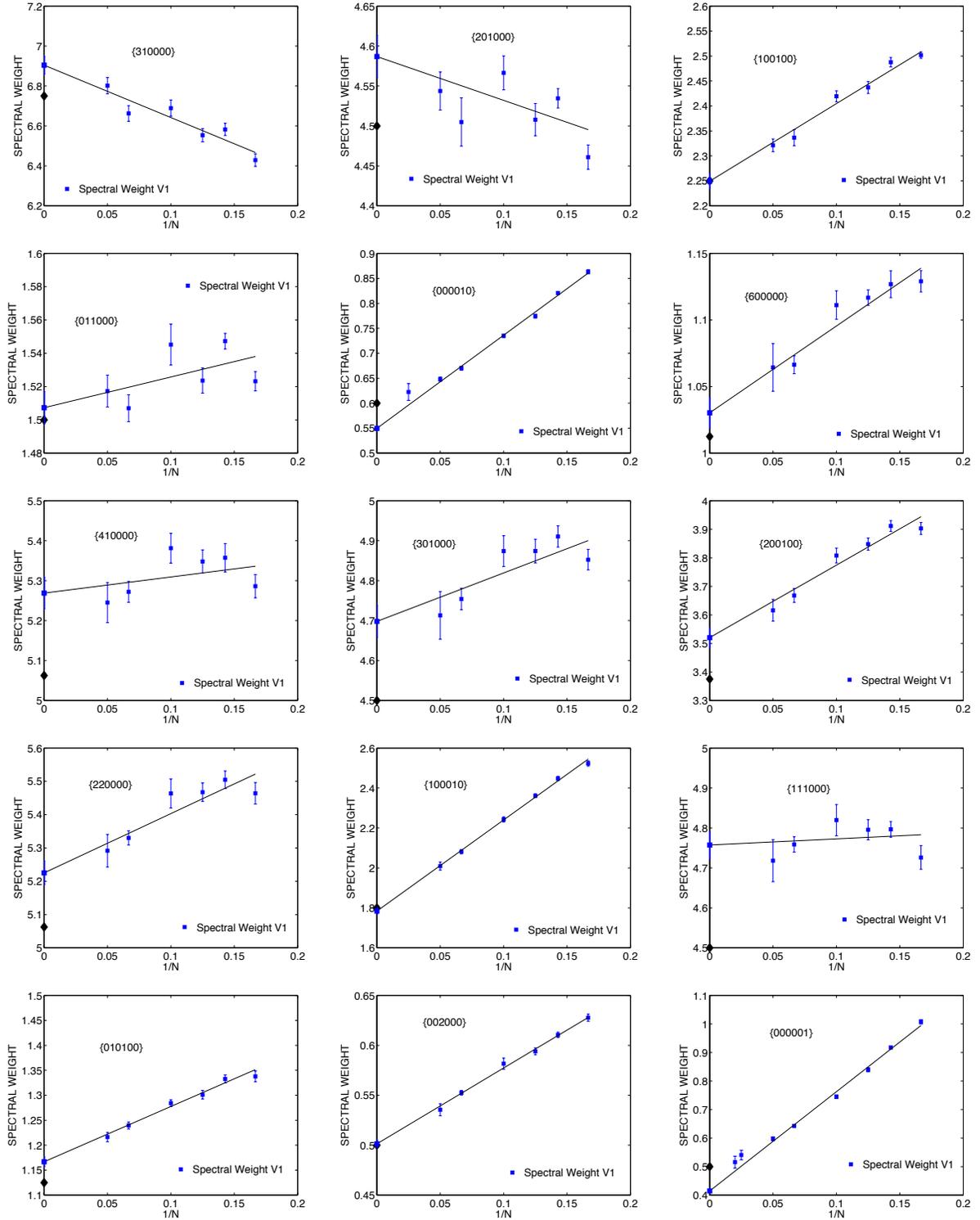}
\caption{(Color online)  Filled squares represent squared spectral weights $|C_{\{n_l\}}|^2$ obtained from Monte Carlo simulaton for $V_1$ interaction for $\Delta M=5 ,6$.  The error bars indicate the statistical uncertainty in the Monte Carlo results. Thermodynamic extrapolation is also shown as a blue square on the vertical axis. Their numerical values are given in Table \ref{table3}.  ECLL  predictions are shown as solid diamonds on the vertical axes.}
\label{MCM5M6}
\end{figure}

\end{widetext}

\clearpage

\appendix

\section{Bosonic operators}

In the standard bosonization approach for one-dimensional Fermi systems, the 
density operators play the role of bosons.  For the state at $\nu=1$, Stone 
observed  [\onlinecite{stone1}] that the fermionic excitations 
at the edge can be mapped into linear combinations of products of symmetric 
polynomials $S_l$'s given in Eq. (\ref{StoneOpDef} ):
\begin{eqnarray}
\label{ExctDet}
\Psi_{\{\lambda_k\}}^N
&=&
\left | \begin{array}{rrrr} z_1^{\lambda _1+N-1} & z_1^{\lambda _2+N-2} & \cdots & z_1^{\lambda _N}  \\
                                          z_2^{\lambda _1+N-1} & z_2^{\lambda_2+N-2} & \cdots  &z_2^{\lambda_N} \\
                                          \vdots & \vdots &\vdots & \vdots\\
					z_N^{\lambda_1+N-1} & z_N^{\lambda_2+N-2} & \cdots &  z_N^{\lambda_N} \\
       \end{array} \right | e^{-\sum_{i=1}^N|z|_i^2/4} \nonumber  \\
 &=& \sum_{\{n_l\}}C^{\{\lambda_k\}}_{\{n_l\}}
 \prod_{l}S_{l}^{n_l} \Phi_{\{0\}}^N,     
  \end{eqnarray}
where $\Phi^N_{\{0\}}$ is the ground state at $\nu=1$, $C^{\{\lambda_k\}}_{\{n_l\}}$'s are the expansion coefficients (cf. Eq 4.12 in [\onlinecite{stone1}]), $\lambda_1\geq \lambda_2\geq \cdots 
\lambda_N$, and $\sum_k \lambda_k=\sum_l l n_l=\Delta M$ is the total angular momentum of the excited state, measured relative to the 
ground state.  The symmetric polynomial 
$S_l$ is identified with bosonic operator $a^\dagger_l$ at angular momentum $l$. 
We now ask under what conditions the operators $S_l$ satisfy the canonical commutation relations.  For this purpose, we use the form given in Eq.(\ref{S_operator}).

The commutation relation $[S_k,S_{l}]=0$, where $k$ and $l$ denote angular 
momentum quantum numbers, follows from the fact that the $S$'s can be expressed  
as polynomials, and can also be verified straightforwardly 
using the operator form of $S$ in Eq. (\ref{S_operator}).  Demonstration of 
\begin{equation}
[S_k,S_l^\dagger]\propto \delta_{kl}
\end{equation}
is more tricky.  An explicit evaluation gives 
\begin{eqnarray}
[S_k,S_l^\dagger] & = & \sum_{m}\sqrt{\frac{(m+k)!(m+l)!}{m!^2} }c_{m+k}^\dagger c_{l+m} \nonumber \\
& & - \sum_{m}\sqrt{\frac{(m+k)!^2}{m!(m+k-l)!}} c_{m+k-l}^\dagger c_{m},
\label{Conjeqn}
\end{eqnarray}
the right hand side of which, in general, does not vanish for $k\neq l$.  

To proceed further, we assume the idealized occupation number 
for the ground state $\Omega$:
\begin{equation}
\langle \Omega| c_k^\dagger c_k|\Omega\rangle   = \left\{  \begin{array}{ll} 
\nu & \textrm{ if $k\le M$} \nonumber \\ 
0 & \textrm{otherwise ,}
\end{array} \right. 
\end{equation}
where $M$ is the angular momentum of the outermost occupied orbital.  This 
is surely correct for $k<<M$, but only approximate near the edge.  We further 
assume that the effect of the operators $S_l$ is confined to the edge; this is 
manifestly correct for $\nu=1$, but not obvious for the FQHE state $\nu=1/m$.

We first consider $k=l$.  Defining normal ordering of operators 
in the usual manner (i.e., by subtracting the ground state expectation value), we get  
\begin{eqnarray}
[S_k,S_k^\dagger]&=&\sum_{m=0}^{\infty}\frac{(m+k)!}{m!}\Big( : c_{m+k}^\dagger c_{m+k}:-:c_{m}^\dagger c_{m}: \nonumber \\
&+&  \langle \Omega| c_{k+m}^\dagger c_{k+m}|\Omega\rangle - \langle \Omega| c_k^\dagger c_k|\Omega\rangle \Big) .
\end{eqnarray}
The dominant contribution comes from the last two terms on the right hand side.
(When only terms near the edge contribute to the commutator,
then for  $m\approx M$,
the factorial term varies as $M^k$, and the normal ordered terms cancel to the 
lowest order.)  This gives 
\begin{eqnarray}
[S_k,S_k^\dagger]&\approx&\sum_{m=0}^{\infty}\frac{(m+k)!}{m!}\left( \langle \Omega| c_{k+m}^\dagger c_{k+m}|\Omega\rangle - \langle \Omega| c_k^\dagger c_k|\Omega\rangle \right) \nonumber \\
&=&  \sum_{m=M-k+1}^{M}  \frac{(m+k)!}{m!}  \left(0-\nu\right) \nonumber \\
&\approx  &-\nu k M^k .
\end{eqnarray}
The canonical bosonic relations are obtained by defining 
\begin{equation}
a_k^\dagger=\frac{1}{\sqrt{\nu k M^k}}S_k, \;\; k>0.
\label{bosondef}
\end{equation}

Next we consider the case when  $k\ne l$, assuming $k>l$ and $N\gg \{k,l\}$ (since only excitations near the edge are significant). The commutator can be expressed as 
\begin{eqnarray}
[S_k,S_l^\dagger]  &\approx& \sum_{m}\Bigg( \sqrt{\frac{(m+k-l)!m!}{(m-l)!^2}} \nonumber \\ &-&\sqrt{\frac{(m+k)!^2}{m!(m+k-l)!}} \Bigg):c_{m+k-l}^\dagger c_m:
 \end{eqnarray}
We have used above that the vacuum expectation values of all combinations on the right hand side vanish due to angular momentum conservation.
Only terms for which $m\approx M $ are significant in the summation.
A little algebra using the asymptotic expansion
\begin{eqnarray}
n^{q-p}\frac{\Gamma(n+p)}{\Gamma(n+q)}&=&1+\frac{(p-q)(p+q-1)}{2n}\nonumber  \\
 && +\frac{1}{12n^2}\displaystyle{p-q\choose 2} \left(3(p+q-1)^2-p+q-1 \right)\nonumber \\
 &&  +\cdots 
\end{eqnarray}
shows that the commutator vanishes to leading order, producing 
\begin{equation}
[S_k,S_l^\dagger]\approx \sum_{m \approx M} \sqrt{m^{k+l}}\mathcal{O}(1/m^2)c_{m+k-l}^\dagger c_m .
\end{equation}
For the bosonic operators defined in Eq.~(\ref{bosondef}), this implies 
\begin{equation}
[a_k^\dagger,a_l]\approx \sum_{m \approx M} \sqrt{\left(\frac{m}{M}\right)^{k+l}} \mathcal{O}(1/m^2)c_{m+k-l}^\dagger c_m \to 0 ,
\end{equation}
thus completing the relationship between $S_l$ and $b_l^\dagger$.
We note that these considerations are valid for arbitrary $\nu$, but 
assume that the effect of the operators $S_l$ is confined to the edge.

\end{document}